# Electron Tunneling Transport and Magnetic Field Modulation through a Superconductor-Kitaev layer-Superconductor Junction

Shiqing Jia[1,2], Yamin Quan[1], Hai-Qing Lin[3,4] * and Liang-Jian Zou[1,2]*

[1] *Key Laboratory of Materials Physics, Institute of Solid State Physics, Chinese Academy of Sciences, P. O. Box 1129, Hefei 230031, China*

[2] *Science Island Branch of Graduate School, University of Science and Technology of China, Hefei 230026, China*

[3] *Beijing Computational Science Research Center, Beijing 100193, China*

[4] *Department of Physics, Beijing Normal University, Beijing, 100875, China*

**Abstract**

We present the electron tunneling transport and its magnetic field modulation of a superconducting (SC) *Josephson* junction with a barrier of single ferromagnetic (FM) Kitaev layer. We find that at $H = 0$, the *Josephson* current $I^S$ displays two peaks at $K/\Delta \approx 3.4$ and $10$, which stem from the resonant tunnelings between the SC gap boundaries and the spinon flat bands and between the SC gap edges and the spinon dispersive bands, respectively. With the increasing magnetic field, $I^S$ gradually decreases and abruptly drops to a platform at the critical magnetic field $h_c = g\mu_B H_c/\Delta \approx 0.03 K/\Delta$, since the applied field suppresses the spinon density of states (DOS) once upon the Kitaev layer enters the polarized FM phase. These results pave a new way to measure the spinon or Majorana fermion DOS of the Kitaev and other spin liquid materials.

*Introduction:* The quantum spin liquid (QSL) phase, which consists of various spin singlet pairings in the spin structure and does not break any constituent symmetries of their underlying lattice, has attracted a great deal of attentions because of its novel nature [1,2]. Enormous efforts have made to understand the essence of the QSLs, especially focusing on the geometrically frustrated interaction with anisotropy [3,4], however, the essence and unique characters of the QSL states remain great debates [5,6]. More than a decade ago Kitaev proposed an exactly solvable model on the two-dimension (2D) honeycomb lattice[7], which shows a ground state of gapless or gapped $Z_2$ QSL with fractionalized excitations [8]. Such a QSL state with gapped excitations has the Abelian anyons [9], the one with gapless excitations may have the non-Abelian anyon excitations [10]. Due to topological protection and large degeneracy of these anyons, the Majorana fermion excitations and its braiding group in the gapless QSL state are believed to be applicable for the quantum computing storage and quantum computation [11,12], hence favorable of the fabrication of the quantum computer. However, how to excite and detect the Majorana fermions in the Kitaev model under magnetic field modulation remains unknown[13-16].

On the other hand, the Josephson tunneling junctions, consisting of two SC leads separated by an insulating or metallic barrier, provide a well probe to measure the quasiparticle information of the central region through the quantum tunneling transport [17,18]. A great deal of central materials, such as insulators [19], normal metals [20], quantum dots [21-30], ferromagnets [31-33] and antiferromagnets [34,35] have been studied. Thus, we construct the SC-Kitaev layer-SC tunneling junctions to study the charge and spin transports of the Cooper pairs, especially the transport of the Majorana fermions, as shown in Fig.1. The SC-Kitaev layer-SC mesoscopic hybrid systems with weak links may open a fruitful research field, not only because of the abundant fundamental features from the interplay between Kitaev physics and SC, but also of the potential application for design and development of new quantum devices.

In this Letter, we would utilize the current feature of the SC-Kitaev layer-SC tunneling junctions to characterize the Majorana fermion modes and its evolutions with increasing magnetic field in the central-zone Kitaev material. These unique behaviors should distinctly differ from the situations with ferromagnetically or antiferromagnetically central barriers. We adopt the non-equilibrium Green's function in the 4×4 Nambu representation [30] to obtain the formula of the normal and

Josephson currents, and find that in the absence of magnetic field, the Josephson current $I^S$ shows two tunneling resonance peaks at K ≈ 3.4Δ and 10Δ, respectively; increasing magnetic field gradually suppresses the current $I^S$ until drops to a small platform at $g\mu_B H_z/\Delta \approx 0.03 K/\Delta$ for FM polarization. One expects that the devices composing of SC and Kitaev materials will contribute more rich and complicated phenomena.

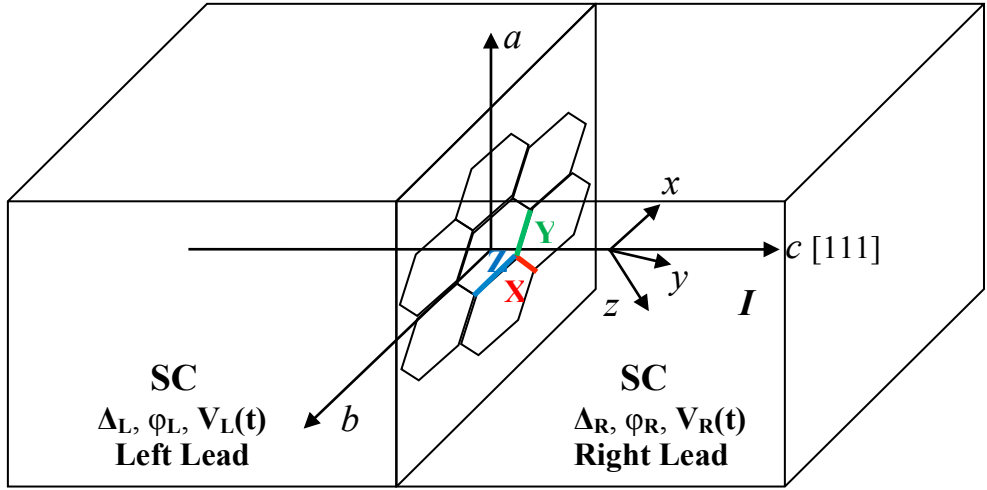

Fig. 1. Schematic superconductor-Kitaev layer-superconductor tunneling junction. The left (right) is the SC lead with gap $\Delta_L$ ($\Delta_R$), phase $\varphi_L$ ($\varphi_R$) and bias votage $V_L(t)$ ($V_R(t)$). The central region is a single-layer Kitaev material in the ab plane with the z-direction ([001]) magnetic field.

*Model Hamiltonian and Formulae:* The total Hamiltonian of the SC-Kitaev QSL-SC tunneling junction under consideration shown in Fig.1. consists of three parts as follows: the left and right SC electrodes $H_{Lead,n}$ ($n$ = L, R), the single-layer Kitaev material in the central scattering region $H_{cen}$, and the interaction part between the SC leads and central material $H_T$, where $S_i^{x(y,z)} = \frac{1}{2}\sum_{i\sigma\sigma'} c_{i\sigma}^\dagger \sigma_{\sigma\sigma'}^{x(y,z)} c_{i\sigma'}$ are the spin components. Let the SC leads be the s-wave superconductors with order parameters $\Delta_n^0 = \Delta_n^0 e^{-i\phi_n}$.

$$H = \sum_{n=L,R} H_{Lead,n} + H_{cen} + H_T(t)$$
$$= \sum_{nk\sigma} \varepsilon_{nk\sigma}^0 a_{nk\sigma}^\dagger a_{nk\sigma} + \sum_{nk}\left[\Delta_n^0 a_{n,-k\downarrow} a_{nk\uparrow} + \Delta_n^0 a_{nk\uparrow}^\dagger a_{n,-k\downarrow}^\dagger\right]$$
$$-K\sum_{<ij>_X} S_i^x S_j^x - K\sum_{<ij>_Y} S_i^y S_j^y - K\sum_{<ij>_Z} S_i^z S_j^z - g\mu_B H_z \sum_i S_i^z \quad (1)$$
$$+\sum_{nk,i,\sigma}\left\{v_{nk,i\sigma}(t)\exp\left[\frac{i\phi_n}{2} + \frac{i}{\hbar}\int_0^t V_n(t_1)dt_1\right]a_{nk\sigma}^\dagger c_{i\sigma} + h.c.\right\}$$

where $a^+_{nk\sigma}$ and $c^+_{i\sigma}$ are the creation operators of electrons in the SC leads and Kitaev layer, respectively; $v_{nk,i\sigma}$ is the hybridization matrix between the SC lead and Kitaev layer, $\varepsilon^0_{nk\sigma}$ is the single-particle energy and $V_n(t)$ is the external electric field; and K is the spin coupling constant in the central Kitaev layer. $H_z$ is the external magnetic field in the $z$-direction (the [001] direction of spin frame) [7].

The current from the $n$th SC lead to the central region

$$I_{n\sigma}(t) = -e\left\langle \frac{dN_{n\sigma}(t)}{dt} \right\rangle = \frac{ie}{\hbar}\left\langle [N_{n\sigma}(t), H(t)]_- \right\rangle$$
$$= \frac{2e}{\hbar}\text{Re}\sum_{ki} v_{nk,i\sigma}(t)\exp\left\{i\left[\frac{\phi_n}{2} + \frac{1}{\hbar}\int_0^t V_n(t_1)dt_1\right]\right\} i\left\langle a^\dagger_{nk\sigma}(t)c_{i\sigma}(t) \right\rangle, \quad (2)$$

is thus a diagonal 4x4 matrix,

$$I_n(t) = \frac{2e}{\hbar}\text{Re}\int dt_1 \left[\mathbf{G}^r(\mathbf{t},\mathbf{t_1})\Sigma^<_{n,ji}(t_1,t) + \mathbf{G}^<(\mathbf{t},\mathbf{t_1})\Sigma^a_{n,ji}(t_1,t)\right]\begin{pmatrix}\sigma_z & \\ & \sigma_z\end{pmatrix}$$
$$= -\frac{2e}{\hbar}\text{Im}\int dt_1 \int \frac{d\varepsilon}{2\pi} e^{-i\varepsilon(t_1-t)/\hbar}\left[f_n(\varepsilon)\rho_n(\varepsilon)\mathbf{G}^r(\mathbf{t},\mathbf{t_1}) + 0.5\rho_n(\varepsilon)\mathbf{G}^<(\mathbf{t},\mathbf{t_1})\right]\mathbf{\Gamma_n}\sum_n\begin{pmatrix}\sigma_z & \\ & \sigma_z\end{pmatrix}. \quad (3)$$

here $\sigma_z$ is the Pauli matrix. The retarded, advanced, and lesser self-energies $\Sigma^{r,a,<}_{n,ji}(t_1,t)$ in the central region are assumed to be independent of the states $i$ and $j$ and $\mathbf{G}^{r,a,<}(\mathbf{t},\mathbf{t_1}) = \sum_{ij} G^{r,a,<}_{ij}(t,t_1)$ are the retarded, advanced, and lesser Green functions, respectively. $\rho_n(\varepsilon)$ is the ratio of the SC density of states $\rho^S_n(\varepsilon)$ to the normal one $\rho^N_n(\varepsilon)$. The Fermi distribution function $f(\varepsilon) = 1/[\exp(\varepsilon/k_B T)+1]$, and the linewidth function is $\Gamma^n_{ji,11}(\varepsilon,t_1,t) \to \Gamma^n_{11}(\varepsilon)$.

We are interested at the SC Josephson current generated by the transport of electron Cooper pairs with zero bias voltage. We can obtain the total SC Josephson current terms for spin-up and spin-down channels,

$$I_{n\uparrow} = -\frac{2e}{\hbar}\text{Im}\int\frac{d\varepsilon}{2\pi}\left\{\begin{array}{l}\Gamma^n_{11}(\varepsilon)\left[f_n(\varepsilon)\rho_n(\varepsilon)\mathbf{G}^r_{11}(\varepsilon) + 0.5\rho_n(\varepsilon)\mathbf{G}^<_{11}(\varepsilon)\right]\\ -\Gamma^n_{21}(\varepsilon)\frac{\Delta^0_n e^{i\phi_n}}{\varepsilon}\left[f_n(\varepsilon)\rho_n(\varepsilon)\mathbf{G}^r_{12}(\varepsilon) + 0.5\rho_n(\varepsilon)\mathbf{G}^<_{12}(\varepsilon)\right]\end{array}\right\}$$

$$I_{n\downarrow} = -\frac{2e}{\hbar}\text{Im}\int\frac{d\varepsilon}{2\pi}\left\{\begin{array}{l}-\Gamma^n_{22}(\varepsilon)\left[f_n(\varepsilon)\rho_n(\varepsilon)\mathbf{G}^r_{22}(\varepsilon) + 0.5\rho_n(\varepsilon)\mathbf{G}^<_{22}(\varepsilon)\right]\\ +\Gamma^n_{12}(\varepsilon)\frac{\Delta^0_n e^{-i\phi_n}}{\varepsilon}\left[f_n(\varepsilon)\rho_n(\varepsilon)\mathbf{G}^r_{21}(\varepsilon) + 0.5\rho_n(\varepsilon)\mathbf{G}^<_{21}(\varepsilon)\right]\end{array}\right\}. \quad (4)$$

Once obtaining the "total" Green functions $\mathbf{G}^{r,a,<}(\varepsilon)$ in the central region with the Dyson equation and Keldish equation, we could get the zero-biased Josephson currents. Throughout this Letter the SC order parameters $\Delta_L$ and $\Delta_R$ in left and right leads have the same modulus/amplitude $|\Delta_L|=|\Delta_R|= \Delta$, but different phase $\varphi=\varphi_L-\varphi_R$. All of the energies are measured in terms of the SC gap amplitude $\Delta$.

*A. Tunneling Process of the SC-Kitaev layer-SC Junction:* In the SC-Kitaev layer-SC Josephson junction shown in Fig.1, the tunneling process of the Cooper pairs depends on the distribution of the DOS of the central Kitaev layer and the SC leads, which could be qualitatively described by the sketched diagram of the unperturbated DOS, $\rho(E)$, shown in Fig. 2. In the DOS plot of the left and right SC leads, the SC energy gaps and phases are $2\Delta_L= 2\Delta$, $\varphi_L$ and $2\Delta_R= 2\Delta$, $\varphi_R$, respectively. In the central Kitaev region, the unperturbated DOS arises from the two kind Majorana fermion modes, one is from the flat bands corresponding to the local Majorana fermion modes, which behave as the δ functions at ±0.1K, respectively; another one is from the linear dispersion bands corresponding to the itinerant Majorana fermion modes, which behave as the approximately linear DOS in the energy range of -0.3K < ε < 0.3K, as seen in Fig.2.

In this Letter we focus on the direct current (DC) in the presence of the SC phase difference φ. The tunneling process of the SC Cooper pairs can be described as follows: The Cooper pair in the left or right lead firstly tunnels into the central Kitaev region, splitting as the quasi-electron and quasi-hole with opposite spins. Due to the strongly correlated insulating and spin liquid nature of the Kitaev region, the spin and charge of the quasi-electron or quasi-hole are separated to form the spinon and holon. Since the thickness of the Kitaev region is only single atomic layer, the tunneling probability of holons is assumed to be unity. In contrast, the Majorana fermion modes display fractional collective excitations, the propagation of the spinons will be modulated by the DOS of the Majorana fermions. Once tunneling out of the central Kitaev region, the separated spinons and holons will recombine to SC Cooper pairs. From the DOS plot and supercurrent formula Eq.(4) one sees that four factors may affect the Josephson current: the amplitude of SC energy gap $\Delta$, the phase difference between the SC left and right leads $\varphi = \varphi_L − \varphi_R$, the linewidth function $\tilde{\Gamma}$, as well as the Kitaev coupling constant K.

Although the spin and charge transports could not be treated completely separately in this study, we separate the spin and charge of quasi-electron or quasi-hole into spinon and holon by employing the slave boson method and assume the tunneling probability of holons to be unity due to the monoatomic Kitaev layer. Thereout, we will realize the tunneling process mainly through the propagation of the spinons or local and itinerant Majorana fermions associated with the Kitaev coupling and modulated by the magnetic field, which is shown in the SC current characteristic.

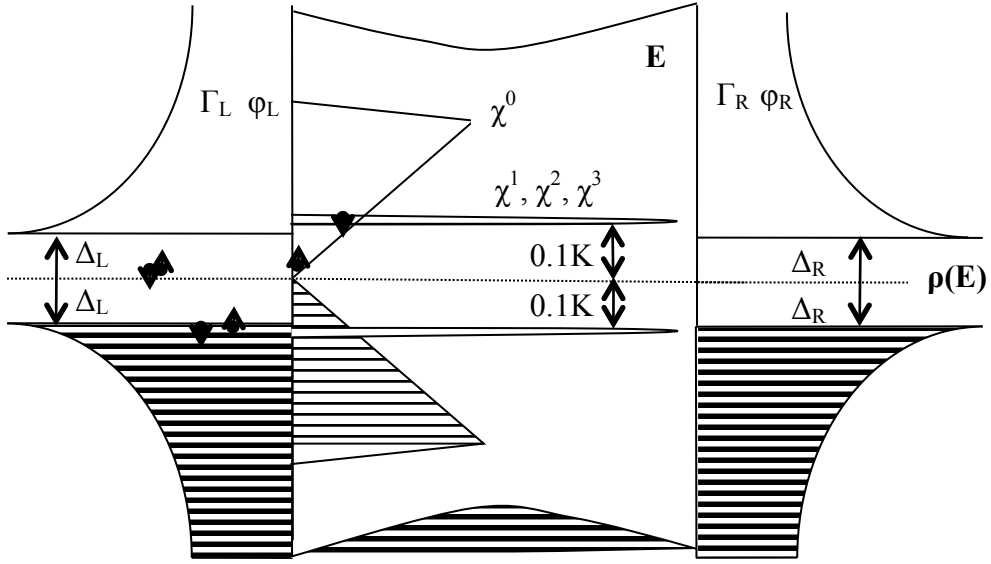

Fig. 2. Plot of the bare density-of-states (DOS) distributions of the superconductor-Kitaev layer-superconductor Josephson junction. The left and the right are the DOS of the left and right SC leads, and the center is that of the Kitaev layer from the spinon bands.

*B. DC Josephson current without magnetic field*: In the absence of external magnetic field, the zero-temperature supercurrent through the Josephson junction as the functions of the Kitaev coupling constant K is shown in Fig. 3(a) for the parameters $\Delta = 1$, $\Gamma = 0.1\Delta$ and $\varphi = 3\pi/2$. In addition, we also plot the energy dependences of the total DOS on the Kitaev couplings in Fig. 3(b).

From Fig.3(a) one finds that there are two obvious peaks at $K = 3.4\Delta$ and $10\Delta$, respectively, in the $I^S$ vs K curves which result from two resonant tunnelings: the former peak comes from the resonant tunneling between the states of the SC gap edge in the left or right lead and the spinon flat band in the central region; the latter from that between the SC gap boundary of the SC leads and the boundary of the spinon dispersive band, as seen in Fig.2. Therefore, we could obtain the information of the spinon flat band and dispersive band corresponding to the local and itinerant

Majorana fermion modes in the central Kitaev material.

Next, when K > 10Δ the tunneling current $I^S$ reaches a maximum, rapidly goes down and becomes negative after K > 12Δ, indicating that the present system transits from a π-junction relation to a general Josephson relation, which stems from the separation of the spinon flat band from the SC gap edge, and the tunneling current of the SC-Kitaev layer-SC junction changes its sign. The equivalent internal "molecular field" characteristic of the spin Kitaev coupling K, namely K<$S_i$>, could have a great influence on the SC phase difference $\varphi$ when K is large enough and finally result in the current sign reversal. With the further increasing K, the opposite DC Josephson current reaches a minimum about K ≈ 13.5Δ and finally gradually approaches zero in extremely large K >= 30Δ.

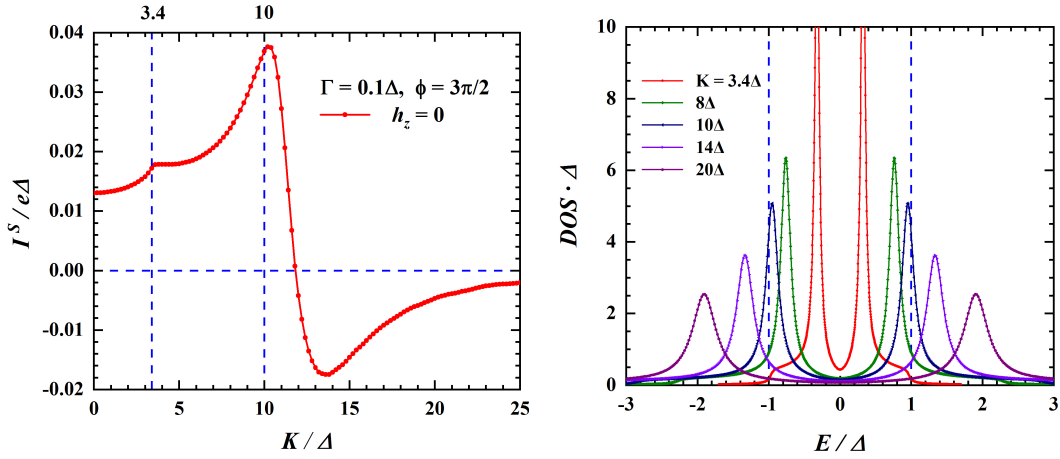

Fig. 3. (a) Kitaev coupling K dependences of the tunneling currents $I^S$ with Δ = 1, Γ = 0.1Δ, and $\varphi$ = 3π/2. (b) Energy E dependences of the total DOS with different Kitaev couplings K. Here $T$ = 0, $h_z$ = 0.

*C. DC Josephson current modulated by magnetic field:* Further, we could self-consistently solve the SC Josephson current in Eq. (4) under the finite external magnetic field. In our earlier work on the Kitaev layer within the Schwinger fermion mean field approach, we have obtained the magnetic phase diagram, and the dependence of the DOS of the spinon energy spectrums $E_{\alpha k\sigma}$ on applied magnetic fields. These results are useful for calculating the magnetic field dependences of the tunneling currents.

The magnetic field dependences of the tunneling currents with different Kitaev couplings K = 2 ~ 20Δ at Γ = 0.1Δ and $\varphi$ = 3π/2 are shown in Fig. 4(a). The particular points $K = 3.4\Delta$ and $10\Delta$ corresponding to two resonant tunneling peaks in the tunneling currents are also plotted. We also plot the magnetic field dependences of the sublattice magnetic moments $m$ of the central Kitaev layer in Fig. 4(b) for different Kitaev couplings at zero temperature.

The tunneling currents $I^S$ decrease almost linearly with increasing magnetic field when $K \leq 10\Delta$ and $h_z = g\mu_B H_z/\Delta < 0.03K/\Delta$. The suppression of the tunneling currents stems from the decrease of the tunneling probability of the suppression of the spinon DOS in the central Kitaev layer by the magnetic field. Further, it is interested that when $K > 10\Delta$, the tunneling current changes to negative sign, indicating the supercurrent reverses the direction. Such a sign reversal in tunneling current arises from the fact that the flat band of the local Majorana fermions lifts higher than the SC gap edge. Meanwhile, the equivalent internal "molecular field" of the spin Kitaev coupling K is large enough to change the phase difference $\varphi$, eventually causing the current to reverse.

However, in the present FM Kitaev junction, the QSL state in the central Kitaev layer is easily broken by external magnetic field, which transits to the polarized FM at $h_c = 0.03K/\Delta$, as seen in Fig. 4(b). And the dispersive spinon spectrums become flat bands and the DOS exhibits two peaks at $\pm(h_z + m)$. When applied magnetic field $h_z$ exceeds the critical field $h_c$, the tunneling current sharply drops to a platform, which is almost independent of $h_z$. These results are in accordance with those of the SC-FM-SC Josephson junction very well [26], as one expects.

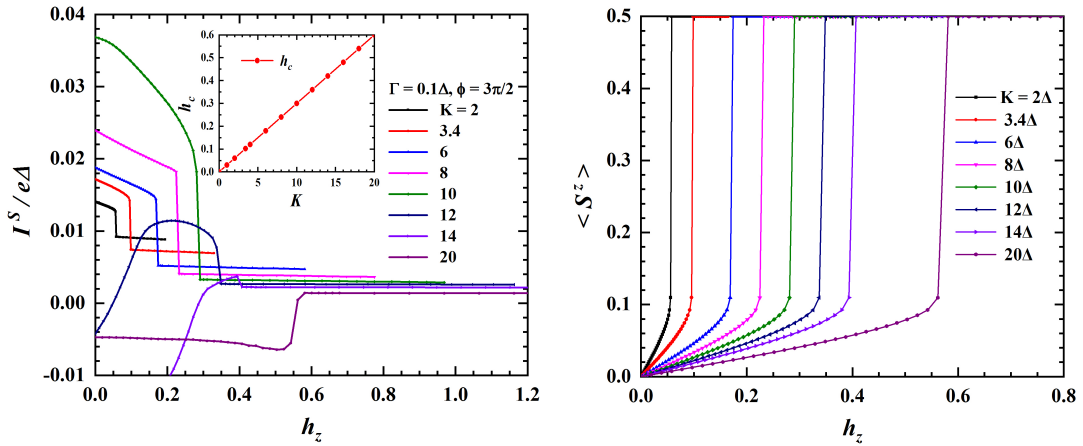

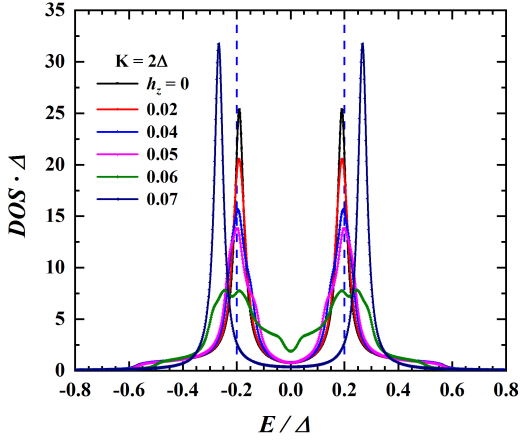

Fig. 4. (a) Magnetic field dependences of the Josephson tunneling currents $I^S$ at $\Gamma = 0.1\Delta$ and $\varphi = 3\pi/2$ for different ferromagnetic Kitaev couplings K. The inset shows the linear dependence of the critical magnetic field $h_c = g\mu_B H_c/\Delta$ on the Kitaev coupling K. (b) Magnetic field dependences of the sublattice magnetic moments $m = <S^z>$ at $T = 0$. (c) Energy E dependences of the total DOS at $K = 2\Delta$ for $h_z = 0 \sim 0.07\Delta$.

Therefore, the SC Josephson current in the SC-single Kitaev layer-SC junction with applied magnetic field is very different from the SC-FM-SC[26] or SC-AFM-SC[29] junctions with single atomic-layer thickness, which go down nearly linearly or parabolically with increasing magnetic field.

*Remarks:* In investigating the electron tunneling transport and its magnetic field modulation in a SC-Kitaev layer-SC *Josephson* junction with the weak link, we assume that the supercurrent majorly depends on the spin tunneling probability of the Cooper pairs. In the absence of magnetic field, due to the resonant tunnelings between the SC gap boundaries and the local Majorana fermion modes and between the SC gap edges and the dispersive itinerant Majorana fermion modes, the *Josephson* current $I^S$ displays two peaks at $K/\Delta \approx 3.4$ and 10, respectively.

With the increasing magnetic field, the applied field suppresses the spinon DOS of the Kitaev layer, the Josephson currents $I^S$ gradually decrease and abruptly drop to a saturation value at the critical magnetic field $h_c = g\mu_B H_c/\Delta \approx 0.03 K/\Delta$ in the polarized FM phase, which distinctly differs from those of the SC-FM-SC [26] and SC-AFM-SC [29] junctions. This may pave a new way to measure the spinon or Majorana fermion DOS of the Kitaev and other spin liquid materials. We expect that our theoretical results could be confirmed by future experiments and be applied in the SC junction devices.

*Acknowledgements:* L. J. thanks the supports from the NSFC of China under Grant Nos. 11774350 and 11474287, H.Q. acknowledges financial support from NSAF U1930402 and NSFC 11734002. Numerical calculations were performed at the Center for Computational Science of CASHIPS and Tianhe II of CSRC.